# Trolljäger:

# Understanding Troll Writing as a Linguistic Phenomenon

Sergei Monakhov, Friedrich Schiller University Jena (Germany)
sergei.monakhov@uni-jena.de

*Abstract*: The current study yielded a number of important findings. We managed to build a neural network that achieved an accuracy score of 91 per cent in classifying troll and genuine tweets. By means of regression analysis, we identified a number of features that make a tweet more susceptible to correct labelling and found that they are inherently present in troll tweets as a special type of discourse. We hypothesised that those features are grounded in the sociolinguistic limitations of troll writing, which can be best described as a combination of two factors: speaking with a purpose and trying to mask the purpose of speaking. Next, we contended that the orthogonal nature of these factors must necessarily result in the skewed distribution of many different language parameters of troll messages. Having chosen as an example distribution of the topics and vocabulary associated with those topics, we showed some very pronounced distributional anomalies, thus confirming our prediction.

# TROLLJÄGER:
# UNDERSTANDING TROLL WRITING AS A LINGUISTIC PHENOMENON

**Introduction**

In February 2018, the U.S. Justice Department indicted 13 Russian nationals associated with the Internet Research Agency (IRA), based in St. Petersburg, for interfering with the 2016 U.S. presidential election (Barrett, Horwitz, & Helderman, 2018). Those individuals were accused of creating false U.S. personas and operating social media pages and groups designed to attract U.S. audiences and to sow discord in the U.S. political system, which included posting derogatory information about a number of candidates, supporting the presidential campaign of then-candidate Donald J. Trump and disparaging Hillary Clinton.

The aftereffects of those events were tremendous: they almost led to the impeachment of the newly elected U.S. president, severely damaged the Russian economy due to the imposed U.S. sanctions, and almost ruined relations between the world's two superpowers. Much more than that, those events revealed and epitomised the ongoing change in modern internet communication, having given prominence to such concepts as 'fake news', 'alternative facts', and even 'post-truth'.

Taking into account the global scale of this scandal and its ever-widening ramifications for human society, one can only wonder why the phenomenon of troll writing has not received, to date, any substantial scientific attention. We do not mean that Russian trolls' tweets were not collected and analysed; they were, as were many other such tweets before them. The problem is that they, to the best of our knowledge, were not viewed as a *language* phenomenon in its own right.

An array of tweet parameters have come under scrutiny: dates of publication, rates of accounts' creation and deletion, language, geographical location, percentage of tweets including images and videos, hashtags, mentions, and URLs; however, hardly any work has been done that could help to explain the discourse-specific features of this type of writing (see Zannettou et al. 2019 for the most recent and comprehensive review). In addition, in the very rare instances of research that has been conducted in the desired direction of content analysis, the findings have been regrettably trivial. For example, in the aforementioned paper, Zannettou et al. discovered that Russian trolls tend to 1) post longer tweets, 2) be more negative and subjective, and 3) refer to more general topics than random Twitter users. Similarly unaspiring from a linguistic perspective are the results of some other recent studies of state-sponsored and inauthentic internet messages (Egele et al. 2017; Elyashar et al. 2017; Volkova & Bell 2016).

The current study aims to fill or rather embark upon filling this gap by addressing the issue on systematic footing, starting from a simple classification task and then diving deeper into analysing the particular language traits of trolls' tweets as opposed to the tweets that we for the lack of better term will refer to as 'genuine'. The structure of the rest of the paper is as follows. In the first section, we describe the process of designing several neural networks that are capable of identifying tweets written by trolls, and we present the obtained results. In the second section, the question of what distinguishes correctly and incorrectly labelled tweets is answered by building a logistic regression model and interpreting its main effects and interactions. In the third section, we resort to topic modelling and analysis of word distribution in trolls' tweets to obtain insights into which parameters characterise troll writing as a linguistic phenomenon. The conclusion synthesises the findings and outlines perspectives for further research.

## What a neural network can predict

Back in November 2017 and in June 2018, the U.S. House Intelligence Committee consecutively released two lists of Twitter handles that had been connected with IRA activity (Permanent Select Committee on Intelligence, 2018), amounting to 3,841 handles in total. On July 25, 2018, Clemson University researchers Darren Linvill, an associate professor of communication, and Patrick Warren, an associate professor of economics, collected every tweet sent from 2,848 of those handles between February 2012 and May 2018, with the vast majority posted from 2015 through 2017 (Linvill & Warren 2018). Linvill and Warren made their dataset, 13 CSV files including 2,973,371 tweets, publicly available on Github, from where we downloaded it.[1]

Choosing a dataset that could be compared with troll tweets turned out not to be a trivial task because a number of assumptions had to be met. First, the overall number of tweets in the second set must be of comparable size to that of a troll's collection of tweets, and such a huge archive is not easy to find. Second, the span of time over which the tweets in the second set were posted must also be somewhere between 2015 and 2017 or near it. Thirdly, the distribution of topics in the second set must be as close as possible to that of trolls' tweets: politics, economy, elections, latest home and foreign affairs. Fourthly, the authenticity of the Twitter accounts in the second set must be beyond any doubt, and they should be as 'un-troll-like' as possible. Finally, and probably most importantly, the tweets of the second subset must be written, as should the trolls' tweets, by a relatively compact group of people allegedly sharing some professional or ideological common ground.

After long consideration we decided upon Tweets of Congress, a collection of daily tweets of both houses of the U.S. Congress, encompassing more than 1,000 campaign, office, committee and party accounts. From this dataset, posted by Alex Litel on Github under permissive MIT License, we obtained 1,468,114 tweets written between June 2017 and June 2019. Given the abundance of requirements, this choice seems to be a reasonable compromise.

The initial stage of work involved data cleaning and preprocessing. From the downloaded archives, we retrieved only the tweets themselves, leaving out all other details such as the author's ID, nickname, date and place of publication, and the number of retweets and comments.

There was an issue with the language of tweets that needed to be addressed: both collections included tweets written in languages other than English. While the trolls' tweets were marked for language, and it was easy to filter relevant messages, the congressmen's tweets contained no such information. Given the size of the collection, we had to resort to automatic language identification. We used the *langid* tool for Python, which has been reported to have better accuracy of identification of microblog messages' language than any of the other systems tested (Lui & Baldwin 2012: 29). The overall classification results were unsatisfactory in our case; however, the problem was solved by filtering out from the tweets classified as not English only those tweets with a confidence score (unnormalised log-probability) of less than 200. In total, 5,054 congressmen's tweets were marked as not written in English and consequently deleted.

We used a *preprocessor* tool that was specially designed for tweet data to remove all unnecessary elements, such as URLs; hashtags; mentions; reserved words (RT, FAV); emojis; and smileys. Punctuation signs and digits were also removed. Then, all tweets were tokenised and lemmatised with the help of a *WordNetLemmatizer* tool based on WordNet lexical database (Fellbaum 1998). After that, the so-called stop words were deleted: all forms of pronouns, auxiliaries and contractions, articles and demonstratives, conjunctions and prepositions. Finally, both collections were checked for duplicate tweets, and all tweets with less than 2 words were disposed of.

---

[1] The data, models, Python and R codes are available at:
https://www.dropbox.com/sh/race05ofob7ga7s/AAAuaBe8Y4p43ViOXdSrRweXa?dl=0

After preprocessing, there remained 1,361,708 trolls' tweets and 1,223,340 congressmen's tweets in our two collections.

For the sake of illustration, this is how one of the trolls' tweets looked before preprocessing:

>   ['Hillary Clinton barks like a dog to make a point about the GOP https://t.co/lnTm0is1VX с помощью @YouTube']

and this is how it looked after:

>   ['hillary', 'clinton', 'bark', 'like', 'dog', 'make', 'point', 'gop']

The next step was preparation of the data for feeding into the neural network, which was chosen as one of the most powerful and up-to-date text classification techniques, surpassing in many respects other non-parametric supervised learning algorithms such as support vector machine (SVM), decision tree, rule induction, and k-nearest neighbours (kNN) (Thangaraj & Sivakami 2018). First, as we did not have enough computational power to work with the whole joint collection of more than 2,500,000 tweets, we randomly selected a sample of 300,000 messages, having previously added 0 at the end of each troll's tweet and 1 at the end of each congressman's tweet—as class labels. The trolls were represented in the sample by 158,564 whereas the congressmen were represented by 141,436 tweets.

After that, all 0s and 1s were removed from the messages and combined into a test vector in the same order as they had been placed in the sample. Next, the test vector was converted into a two-dimensional Numpy array to serve as a reference tool for the neural network so that each 0 became [1, 0], and each 1 became [0, 1].

Since neural networks cannot work with strings of text but only with numbers, all tweets had to be encoded. Vectorising was performed with the help of the *Tokenizer* preprocessing tool (Chollet 2015): each lemma in the sample was assigned a unique index, and then each tweet was turned into a sequence of integers, each integer being the index of a lemma in the dictionary.

All input vectors fed into a neural network must be of the same length; therefore, all encoded tweets were padded to the same length: the length of the longest message in our sample (73 words). Every tweet shorter than that was padded by adding the sufficient number of 0s at the end.

Finally, the encoded tweets were split into two random subsets: one of 240,000 messages for training a neural network and another of 60,000 messages for testing the accuracy of classification (Pedregosa et al. 2011).

For our classification purposes, we compared two neural networks: the simplest feedforward sequential model, that is, in essence, just a linear stack of layers, and a much more advanced bidirectional long short term memory model (BLSTM) that duplicates the first recurrent layer in the network and then provides the first layer with the input sequence 'as is' and the second layer with the reversed copy of the input (Gers et al. 2002; Baldi et. al. 2001; Hochreiter & Schmidhuber 1997; Schuster & Paliwal 1997). As it had been reported that BLSTMs are the fastest and most accurate among other recurrent neural networks (Graves & Schmidhuber 2005), we propose that our second model should outperform the first one.

The configuration of two neural networks is presented in Figures 1 and 2, respectively. The sequential model includes the following: 1) one embedding layer that turns the word indexes of each tweet into dense vectors of a fixed size (73 is the number of words in the longest tweet, 100 is the number of neurones in each layer of the network); 2) two hidden fully connected layers; and

3) an output layer with two possible outcomes: 0 or 1. In the BLSTM model, the embedded and output layers are the same; however, instead of two hidden layers, it includes the following: 1) a bidirectional layer (that is, two duplicated layers for the proper and reversed input, hence 200 instead of 100 neurones); 2) a time-distributed layer that applies the following dense layer to every temporal slice of an input; and 3) the fully connected layer itself. The first number of trainable parameters in both models, 800970, is calculated via multiplication of the number of inputs (80096 words in vocabulary + 1) and the number of outputs (100 units in the hidden layer).

Figure 1. Topology of the sequential network

```
Layer (type)                    Output Shape              Param #
=================================================================
input_2 (InputLayer)            (None, 73)                0
_________________________________________________________________
embedding_5 (Embedding)         (None, 73, 100)           8009700
_________________________________________________________________
bidirectional_2 (Bidirection    (None, 73, 200)           160800
_________________________________________________________________
time_distributed_2 (TimeDist    (None, 73, 100)           20100
_________________________________________________________________
flatten_5 (Flatten)             (None, 7300)              0
_________________________________________________________________
dense_14 (Dense)                (None, 100)               730100
_________________________________________________________________
dense_15 (Dense)                (None, 2)                 202
=================================================================
Total params: 8,920,902
Trainable params: 8,920,902
Non-trainable params: 0
```

Figure 2. Topology of the BLSTM network

```
Layer (type)                    Output Shape              Param #
=================================================================
embedding_3 (Embedding)         (None, 73, 100)           8009700
_________________________________________________________________
dense_7 (Dense)                 (None, 73, 100)           10100
_________________________________________________________________
dense_8 (Dense)                 (None, 73, 100)           10100
_________________________________________________________________
flatten_3 (Flatten)             (None, 7300)              0
_________________________________________________________________
dense_9 (Dense)                 (None, 2)                 14602
=================================================================
Total params: 8,044,502
Trainable params: 8,044,502
Non-trainable params: 0
```

In compiling the models, we used a rectified linear unit ('relu') as an activation function for each hidden layer; this type of neurone propagates forward all positive inputs unchanged but returns 0s for all negative inputs (Hahnloser, Seung, & Slotine 2003). For the output layer, softmax activation was used; this function turns the scores from the last hidden layer into probabilities that sum up to 1, thus allowing the model to classify each tweet as 0 (written by trolls) or 1 (written by congressmen). Categorical crossentropy was chosen as the loss function, and the stochastic gradient descent was optimised by an 'adam' algorithm (Kingma & Ba 2015). Each model was trained on 168,000 samples and validated on 72,000 samples in 10 epochs.

Table 1. Neural networks' loss and accuracy scores (training and validation sets)

|  | Epochs | Time | Training | | Validation | |
| --- | --- | --- | --- | --- | --- | --- |
|  |  |  | Loss | Accuracy | Loss | Accuracy |
| *Sequential network* | 1 | 354s | 0.2498 | 0.8996 | 0.2302 | 0.9086 |
|  | 2 | 349s | 0.1810 | 0.9262 | 0.2382 | 0.9098 |
|  | 3 | 403s | 0.1388 | 0.9393 | 0.2704 | 0.9022 |
|  | 4 | 410s | 0.0996 | 0.9504 | 0.3309 | 0.8925 |
|  | 5 | 426s | 0.0708 | 0.9593 | 0.3916 | 0.8826 |
|  | 6 | 419s | 0.0533 | 0.9659 | 0.4885 | 0.8803 |
|  | 7 | 424s | 0.0429 | 0.9721 | 0.5647 | 0.8741 |
|  | 8 | 421s | 0.0349 | 0.9754 | 0.6528 | 0.8679 |
|  | 9 | 422s | 0.0295 | 0.9797 | 0.7479 | 0.8702 |
|  | 10 | 426s | 0.0258 | 0.9828 | 0.8163 | 0.8659 |
| *BLSTM* | 1 | 1547s | 0.2496 | 0.8996 | 0.2236 | 0.9118 |
|  | 2 | 1327s | 0.1874 | 0.9262 | 0.2282 | 0.9067 |
|  | 3 | 1342s | 0.1536 | 0.9393 | 0.2483 | 0.9076 |
|  | 4 | 905s | 0.1254 | 0.9504 | 0.2643 | 0.9030 |
|  | 5 | 897s | 0.1020 | 0.9593 | 0.3034 | 0.8981 |
|  | 6 | 894s | 0.0837 | 0.9659 | 0.3691 | 0.8979 |
|  | 7 | 908s | 0.0693 | 0.9721 | 0.4676 | 0.8940 |
|  | 8 | 890s | 0.0593 | 0.9754 | 0.4738 | 0.8935 |
|  | 9 | 894s | 0.0499 | 0.9797 | 0.4999 | 0.8952 |
|  | 10 | 896s | 0.0420 | 0.9828 | 0.5775 | 0.8881 |

The obtained scores for both models are given in Table 1. From the validation scores, it is clear that both models began overfitting from the second epoch: while the accuracy of the classifying tweets from the training set rose steadily, the accuracy of the classifying tweets from the validation moved in the opposite direction. This result was confirmed by testing both models on 60,000 tweets from the test set.

The overall results of classification are very satisfactory: such accuracy scores mean that out of 100 random new tweets, our neural networks can correctly classify 87 and 89 messages, respectively. In addition, we can see that, as expected, the BLTSM outperformed the sequential network, although in terms of computational costs, its work was much more expensive.

Can these results be improved? To answer the question, we built another neural network that resembled our BLTSM model in all respects but one: instead of the random initialisation of weights in the network, we used pretrained Word2Vec embeddings, a very efficient and popular algorithm for representing words as continuous vectors comprising the probabilities of their co-occurrences with every other single word in a given corpus (Mikolov et al. 2013). As Kocmi &. Bojar argue, 'pretrained embeddings, as opposed to random initialisation, could work better because they already contain some information about word relations' (2017).

We decided not to download and use ready-made 'generic' Word2Vec embeddings that are freely available on the internet[2] but rather build our own Word2Vec model with the aim of trying to catch specific connections between the words in our collection of tweets. The hyperparameters of the model were as follows: the size of the context window equal to 3, the number of iterations equal to 5, the minimal occurrence threshold equal to 5. Importantly, the model was trained on the whole collection of 2,585,048 tweets, not only on the sample that was randomly selected for the neural network.

After that, an embedding matrix was created. For all words in the tweets of our sample that had been vectorised by the Word2Vec model, the matrix was filled with respective vectors (hence there were 100 neurones in the network: the Word2Vec vectors that we used were 100-dimensional). For the words that had not been vectorised, the matrix was filled with zeroes.

Pretrained embeddings are sometimes used as a fixed mapping for a more robust topology (Kenter and De Rijke 2015); however, we preferred to leave them trainable as all other weights in the network. A new neural network was, again, trained, validated and tested. Its loss and accuracy scores are provided in Table 2 for ease of comparison alongside the already cited results of the two previous models. The improvement was highly significant. We have little doubt that with further fine-tuning of the network the results can improve even more; however, for now, we will leave it as it is.

Table 2. Loss and accuracy scores of models without pretrained embeddings and with them

|  | Loss | Accuracy |
|---|---|---|
| *Sequential network* | 0.81 | 0.87 |
| *BLSTM* | 0.56 | 0.89 |
| *BLSTM with pretrained embeddings* | 0.33 | 0.91 |

Russian trolls were engaged in swaying the U.S. elections in favour of Donald Trump, who himself is infamous for provocative tweeting. It is reasonable to expect that if our neural network can generalise very well, even across completely different samples, it would classify the majority of the POTUS tweets as being troll-like. To test this hypothesis, we downloaded and analysed 32,804 tweets posted by Trump from 2009 to 2017. Those tweets were preprocessed in exactly the same way that trolls' and the congressmen's tweets had been, and after that, the whole sample was fed into our BLSTM neural network with pretrained embeddings for classification.

---

[2] E.g., here: https://code.google.com/archive/p/word2vec/.

The results were unsurprising: the model classified 22,055 tweets as troll-like and only 10,749 as 'genuine'. Therefore, we could conclude that each time Trump posts a tweet, there is a 67 % probability that it would be something that a paid internet troll could have written.[3]

However, what exactly does it mean to be troll-like from a linguistic perspective? We will try to answer this question in the following sections.

**What a neural network cannot predict**

It is notoriously difficult to determine what exactly occurs inside a neural network; however, it is of paramount importance to us to identify some of the characteristic traits of the tweets that our model has classified correctly and those for which it failed. Only by doing this can we hope to obtain some insight into what actually distinguishes troll writing from genuine writing on the language level.

First, we divided our *test_predictions_tweets* data (60,000 tweets by both trolls and congressmen on which the model was tested) into *right_tweets* and *wrong_tweets* subsets—for the precisely and mistakenly classified tweets, respectively. Then, we made the dictionaries of all the lexemes in both subsets and calculated the number of words and lemmas as well as the type/token ratios. The results are provided in Table 3.

Table 3. TTR of correctly and incorrectly classified tweets

|  | Number of words | Number of lemmas | TTR (%) |
|---|---|---|---|
| *right_tweets* | 646,092 | 33,908 | 5.24 |
| *wrong_tweets* | 48,135 | 9,522 | 19.78 |
| *test_predictions_tweets* | 694,227 | 35,362 | 5.09 |

The surprising aspect is the TTR for the *wrong_tweets* subset: it is more than 3 times larger than the TTR of the *right_tweets* subset and the TTR of the whole sample. Since it is assumed that a high TTR indicates a high degree of lexical variation while a low TTR indicates the opposite (Templin 1957), one could guess that the reason behind the problems that our neural network experienced with *wrong_tweets* is their higher degree of lexical diversity: there are many words that are used just a few times.

Table 4. Ratios of unique and shared lemmas in correctly and incorrectly classified tweets

|  | Number of lemmas | Unique lemmas | Ratio (%) | Shared lemmas | Ratio (%) |
|---|---|---|---|---|---|
| *right_tweets* | 33,908 | 25,840 | 76.20 | 8,068 | 23.79 |
| *wrong_tweets* | 9,522 | 1,454 | 15.26 | 8,068 | 84.73 |

To check this assumption, we calculated the number of the lemmas that are unique for the *wrong_tweets* subset, the *right_tweets* subset, and those that are used in both subsets. The picture (Table 4) is quite the opposite of what one would expect. While the *right_tweets* are characterised by the enormous ratio of lemmas that are used exclusively within this subset, the situation is reversed for the *wrong_tweets*: almost 85 % of all their lemmas are those that are also encountered in another subset.

---

[3] Unfortunately, Trump's tweets were not chronologically arranged in the collection that we used. Given a large span of covered time, it would be very interesting to compare the tweets posted before Trump began his election campaign and those posted after.

Pearson's Chi-squared test with Yates' continuity correction lends credibility to these findings: X-squared = 3520.5, df = 1, p-value < 2.2e-16. As Figure 3 clearly illustrates, while the *wrong_tweets* exhibit greater lexical diversity, it is in the *right_tweets* subset where the unique lemmas are significantly overrepresented.

Figure 3. Association plot of the representation of unique lemmas
in correctly and incorrectly classified tweets

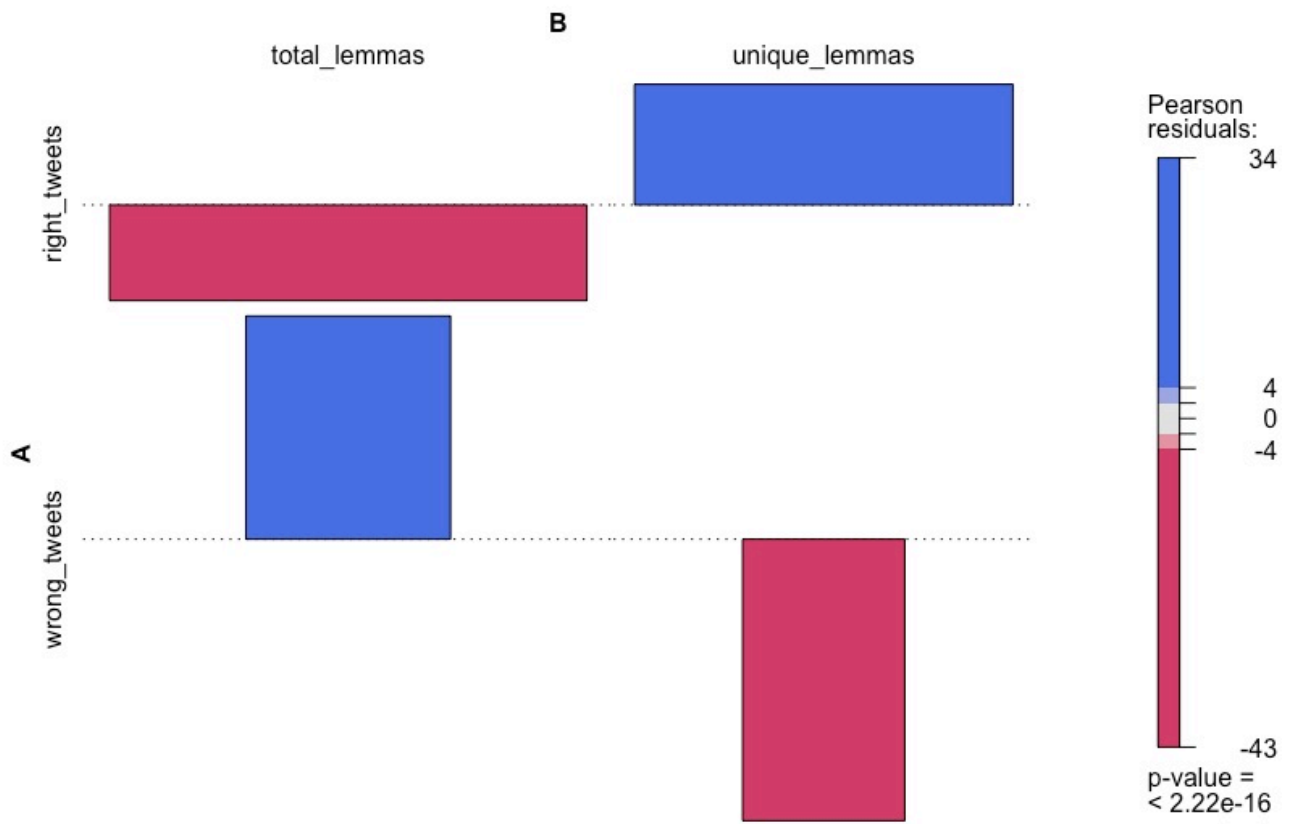

Nonetheless, analysis of the unique lemmas distribution in the tweets of both subsets, the results of which are provided in Table 5, proves that those lemmas alone cannot account for the discrepancies in the text classification. More than 30 % of all *right_tweets* and more than 70 % of all *wrong_tweets* do not contain any unique lemmas at all.

Table 5. Ratios of correctly and incorrectly classified tweets
with unique and shared lemmas

|  | Number of tweets | With shared lemmas | Ratio (%) | With unique lemmas | Ratio (%) |
| --- | --- | --- | --- | --- | --- |
| *right_tweets* | 54,650 | 54,441 | 99.61 | 35,922 | 65.73 |
| *wrong_tweets* | 5,350 | 5,347 | 99.94 | 1,143 | 21.36 |

Therefore, our next goal was to identify a list of variables and build a regression model that could predict which tweet in a given corpus has higher odds of being classified correctly by a neural network.

In determining the variables, it makes sense to take into account only those parameters that could influence the neural network's decisions, that is, only distributional information is relevant while all semantic and morphosyntactic variation may be excluded from consideration. The following list of variables was finally identified:

(1) mean pairwise cosine distance within tweet (**cm**),
(2) standard deviation of cosine distances within tweet (**csd**);
(3) range of cosine distances within tweet (**cr**);
(4) mean word frequency (**mf**);
(5) relative length of tweet (**rl**);
(6) unique words ratio per tweet (**uwr**);
(7) cross-tweet words representation (**ctwr**).

The variables (1)—(3) are fairly much self-explanatory. To calculate the numbers, we used vector representations of the most frequent words of the tweets provided by our Word2Vec model. That required, first, building the vocabulary of the Word2Vec model, removing from the tweets for all the words that were not vectorised by the model and dividing the rest into all possible pairings so that for a tweet consisting of 14 words

['past', 'time', 'stop', 'turn', 'blind', 'eye', 'deaf', 'ear', 'children', 'wait', 'bear', 'time', 'end', 'suffer']

we would obtain 90 pairs of words:

[['past', 'time'], ['past', 'stop'], ['past', 'turn'], ['past', 'blind'], ['past', 'eye'], ['past', 'deaf'], ['past', 'ear'], ['past', 'children'], ['past', 'wait'], ['past', 'bear'], ['past', 'time'], ['past', 'end'], ['past', 'suffer'], ['time', 'stop'], ['time', 'turn'], ['time', 'blind'], ['time', 'eye'], ['time', 'deaf'], ['time', 'ear'], ['time', 'children'], ['time', 'wait'], ['time', 'bear'], ['time', 'end'], ['time', 'suffer'], ['stop', 'turn'], ['stop', 'blind'], ['stop', 'eye'], ['stop', 'deaf'], ['stop', 'ear'], ['stop', 'children'], ['stop', 'wait'], ['stop', 'bear'], ['stop', 'time'], ['stop', 'end'], ['stop', 'suffer'], ['turn', 'blind'], ['turn', 'eye'], ['turn', 'deaf'], ['turn', 'ear'], ['turn', 'children'], ['turn', 'wait'], ['turn', 'bear'], ['turn', 'time'], ['turn', 'end'], ['turn', 'suffer'], ['blind', 'eye'], ['blind', 'deaf'], ['blind', 'ear'], ['blind', 'children'], ['blind', 'wait'], ['blind', 'bear'], ['blind', 'time'], ['blind', 'end'], ['blind', 'suffer'], ['eye', 'deaf'], ['eye', 'ear'], ['eye', 'children'], ['eye', 'wait'], ['eye', 'bear'], ['eye', 'time'], ['eye', 'end'], ['eye', 'suffer'], ['deaf', 'ear'], ['deaf', 'children'], ['deaf', 'wait'], ['deaf', 'bear'], ['deaf', 'time'], ['deaf', 'end'], ['deaf', 'suffer'], ['ear', 'children'], ['ear', 'wait'], ['ear', 'bear'], ['ear', 'time'], ['ear', 'end'], ['ear', 'suffer'], ['children', 'wait'], ['children', 'bear'], ['children', 'time'], ['children', 'end'], ['children', 'suffer'], ['wait', 'bear'], ['wait', 'time'], ['wait', 'end'], ['wait', 'suffer'], ['bear', 'time'], ['bear', 'end'], ['bear', 'suffer'], ['time', 'end'], ['time', 'suffer'], ['end', 'suffer']] and, after measuring the cosine distances between the vector representations in each pair, the following 90 numbers:

[0.23084405, 0.165006, 0.09426032, 0.018875737, 0.08728446, -0.035145633, -0.21513806, 0.15835008, 0.39891496, 0.031234484, 0.23084405, 0.42888537, 0.15516211, 0.3253909, 0.23327011, -0.052872457, 0.049503084, -0.10375977, -0.15152891, 0.2545478, 0.4174173, 0.014464399, 0.29235175, 0.13149995, 0.31327742, 0.12590031, 0.0901772, -0.06048733, -0.048911296, 0.20062421, 0.22691028, -0.08247886, 0.3253909, 0.6127991, 0.105595976, 0.3352162, 0.18447617, 0.12916106, 0.062038686, 0.15504947, 0.3920124, 0.07559819, 0.23327011, 0.23965108, 0.18687809, 0.45571902, 0.43940967, 0.2538958, 0.16856779, 0.13967085, 0.1353369, -0.052872457, 0.043003216, 0.14657624, 0.30956715, 0.3490589, 0.16273755, 0.17156608, 0.104764484, 0.049503084, 0.039942108, 0.061188716, 0.461818, 0.14407241, 0.057416767, 0.14238377, -0.10375977, -0.09211613, 0.11603537, -0.023796022, 0.03380314, 0.15804541, -0.15152891, -0.24166468, 0.031274837, 0.17082833, 0.16026187, 0.2545478, 0.26816827, 0.32470933,

0.052134376, 0.4174173, 0.24416408, 0.22999296, 0.014464399, 0.046255637, 0.23251586, 0.29235175, 0.13149995, 0.23462225]

from which it follows that the most closely related words in our example are *stop* and *end* (0.6127991)[4] while the most distant ones are *ear* and *end* (-0.24166468). After that, we only had to calculate the mean, standard deviation, and range for all cosine similarities. In our example, that makes [0.15050405], [0.16194732], and [0.37113443], respectively.

Variable (4), mean word frequency, was constructed by replacing all the words in tweets with their respective frequencies in the whole sample and calculating the mean. The aforementioned tweet after this procedure turned into the following sequence of numbers:

[315, 2432, 1186, 398, 42, 145, 11, 9, 866, 298, 215, 2432, 1038, 152],

since the words *past*, *time*, *stop*, *turn*, *blind*, *eye*, *deaf*, *ear*, *children*, *wait*, *bear*, *time* (appear for the second time), *end*, and *suffer* are used throughout the sample the respective number of times. The mean word frequency for this tweet equals 681.35.

Variable (5), the relative length of a tweet, gives the number of words within tweet divided by the mean length of all tweets in the sample. As the latter measure is 11.57, the relative length of our example with its 14 words will be equal to 1.20.

Variable (6), the measure of the unique words ratio per tweet, is, in fact, another way of calculating the mean word frequency but additionally taking into account the distribution of the lemmas. The *test_predictions_tweets* sample was vectorised by turning each tweet into a sequence of integers (each integer being the index of a lemma in a dictionary). For each integer, the respective word count was obtained.

If we consider the array of unique indices of all the lemmas in the sample (1) and the array of their respective frequencies (2), it will become clear that the lemmas are indexed not alphabetically or chronologically (in order of appearance) but with regard to their total number of counts:

(1) [1, 2, 3, ..., 47610, 47611, 47612]
(2) [4766, 3501, 3318, ..., 1, 1, 1]

so that the word with index 1 appears 4,766 times, the word with index 2—3,501 times, and the last words with indices from 47610 to 47612 are hapax legomena. The first 7 entries of the resulting word-index dictionary are given below, with the word Trump being, not surprisingly, number 1, an absolute leader: {'trump': 1, 'today': 2, 'people': 3, 'woman': 4, 'work': 5, 'make': 6, 'American': 7, ...}.

Thus, the unique words ratio was obtained by calculating the mean of indices of all words within a single tweet. It goes without saying that the greater this value, the more bottom-ranking words of the dictionary will be used in a tweet. In our example, it is 1415.64—which is quite different from the respective mean word frequency (681.35).

Variable (7), the measure of cross-tweet words representation, is, perhaps, the most difficult to grasp. To avoid confusion, below, we will include the relevant fragment of our Python code and comment on it in more detail.

```
(1)
signal_lemmas_wrong = [lemma for lemma in test_tweets_lemmas \
        if lemma in wrong_tweets_lemmas if lemma not in right_tweets_lemmas]
signal_lemmas_right = [lemma for lemma in test_tweets_lemmas \
        if lemma not in wrong_tweets_lemmas if lemma in right_tweets_lemmas]
```

---

[4] Cosine measure returns similarities in the range <-1, 1> (the greater, the more similar).

In step (1), we make two lists of lemmas that are used exclusively in *right_tweets* and *wrong_tweets*.

(2)
```
wrong_word_gradience = []
for tweet in wrong_tweets:
   for word in tweet:
      if word in signal_lemmas_both:
         wrong_word_gradience.append(word)

unique_wwg, counts_wwg = np.unique(wrong_word_gradience, return_counts=True)
from scipy import stats
counts_wwg = stats.zscore(counts_wwg)
dictionary_wwg = dict(zip(unique_wwg, counts_wwg))

right_word_gradience = []
for tweet in right_tweets:
   for word in tweet:
      if word in signal_lemmas_both:
         right_word_gradience.append(word)

unique_rwg, counts_rwg = np.unique(right_word_gradience, return_counts=True)
counts_rwg = stats.zscore(counts_rwg)
dictionary_rwg = dict(zip(unique_rwg, counts_rwg))

dict_both = [dictionary_rwg, dictionary_wwg]
dict_gradience = {}
for k in dictionary_rwg.keys():
   dict_gradience[k] = tuple(dict_gradience[k] for dict_gradience in dict_both)

keys_dict_gradience = list(dict_gradience.keys())
values_dict_gradience = list(dict_gradience.values())

signal_lemmas_right_extend = []
for i in range(len(values_dict_gradience)):
   if values_dict_gradience[i][0] > values_dict_gradience[i][1]:
      signal_lemmas_right_extend.append(keys_dict_gradience[i])

signal_lemmas_wrong_extend = []
for i in range(len(values_dict_gradience)):
   if values_dict_gradience[i][0] < values_dict_gradience[i][1]:
      signal_lemmas_wrong_extend.append(keys_dict_gradience[i])
```

In step (2), two supplementary lists are created: one of the lemmas that co-appear more often with lemmas used exclusively in *wrong_tweets*; another of the lemmas that co-appear more often with lemmas used exclusively in *right_tweets*. The obtained scores were normalised. The distribution of words in our example is given in Table 6.

Table 6. Cross-tweet words representation measure exemplified by one tweet

|      | right_tweets | wrong_tweets |
|-----:|:------------:|:------------:|
| *past* | 1.12 | 0.92 |
| *time* | 10.85 | 12.21 |
| *stop* | 5.11 | 5.78 |
| *turn* | 1.43 | 2.28 |
| *blind* | -0.19 | -0.24 |
| *eye* | 0.40 | 0.46 |
| *deaf* | — | — |
| *ear* | -0.33 | -0.18 |
| *children* | 3.77 | 2.35 |
| *wait* | 0.99 | 1.44 |
| *bear* | 0.73 | 0.59 |
| *time* | 10.85 | 12.21 |
| *end* | 4.60 | 2.67 |
| *suffer* | 0.38 | 0.79 |

It is evident that *time* and *suffer* are more strongly associated with unique lemmas of the *wrong_tweets* subset whereas *children* and *end* appear more frequently together with lemmas that are characteristic for *right_tweets*.

```
(3)
prob1 = [0 for i in test_predictions_tweets]
prob2 = [0 for i in test_predictions_tweets]
k = 0

for i in test_predictions_tweets:
   for j in i:
      if j in signal_lemmas_right or j in signal_lemmas_wrong:
         prob1[k] += 1.1
         prob2[k] += 0.1
      else:
         if j in signal_lemmas_right_extend or j in signal_lemmas_wrong_extend:
            prob1[k] += 0.5
            prob2[k] += 0.1
         else:
            prob1[k] += 0.1
            prob2[k] += 1.1

   k += 1

prob_reg = [prob1[i]/prob2[i] for i in range(len(test_predictions_tweets))]
```

Most of the heavy lifting is done in step (3). To begin with, for each lemma in the sample two zero values are assigned: *prob1* and *prob2*, which are numerical indications of the 'biasedness' and the 'unbiasedness' of this word, respectively. The algorithm then takes each tweet in the sample and each word in the tweet, consecutively, and first checks whether the word belongs to the short or to the extended list of lemmas that are attracted to *right_tweets* or *wrong_tweets*.

If yes, the *prob1* of the word in question is enlarged by 1.1 (for core lemmas) or 0.5 (for associated lemmas) while the *prob2* is enlarged by 0.1 for both types (to avoid zeroes in the array). If that is not the case, then the word's *prob1* is enlarged by 0.1 and its *prob2* is enlarged by 1.1. Finally, the measure of cross-tweet words representation within a tweet is created by calculating the ratio of the sum of the *prob1* values and the sum of the *prob2* values for each tweet.

Returning to our example, we obtain for it *prob1* = 7.6 (0.5 x 13 + 1.1 x 1) and *prob2* = 1.4 (0.1 x 14): all the words in this tweet belong to extended subsets; thus, they obtain *prob1* = 0.5 with the only exception of the word *deaf*, which appears only in *right_tweets*. This results in a *prob_reg* = 7.6/1.4 = 5.42.

A summary of how the values of each variable should be interpreted can be found in Table 7. Finally, our categorical response variable (**pr**) includes 2 possible outcomes: 0 for the tweets that were misclassified and 1 for the tweets that were labelled correctly.

Table 7. Interpretation of variables in regression analysis

| Variable | Interpretation: the greater a value... |
| --- | --- |
| *mean pairwise cosine distance within tweet (cm)* | ...the more often words of a tweet appear together in other tweets |
| *standard deviation of cosine distances within tweet (csd)* | ...the more diverse are words of a tweet in terms of their cooccurrence |
| *range of cosine distances within tweet (cr)* | ...the greater a distance between the most similar and most unnatural pairs of words in a tweet |
| *mean word frequency (mf)* | ...the more frequently words of a tweet are used in the whole corpus |
| *relative length of tweet (rl)* | ...the more a tweet surpasses the average length in the whole corpus |
| *unique words ratio per tweet (uwr)* | ...the more bottom-ranking words of the whole corpus's vocabulary are used in a tweet |
| *cross-tweet words representation (ctwr)* | ...the more words with a limited distribution are used in a tweet |

As *right_tweets* outnumber *wrong_tweets* by a huge margin (54,650 and 5,350 tweets, respectively), we randomly selected 4,000 tweets from each subset and merged them so that our data frame for regression analysis became perfectly balanced.

To determine which of the 7 independent variables influence the outcome of the classification task the most, we fitted a binary logistic regression model to the data, with (cm) mean pairwise cosine distances, (csd) standard deviation of all cosine distances, (cr) range of cosine similarities, (mf) mean word frequency, (rl) relative length of tweet, (uwr) unique words ratio, and (ctwr) cross-tweet words representation as main effects; no interactions were taken into account in this **Model.1**. Overall, Model.1 was highly significant: LR chi2(7)= 4024.4, p < 0.001 and explained a reasonable amount of variance: pseudo-$R^2$ = 0.36 (McFadden), 0.39 (Cox and Snell), 0.52 (Nagelkerke).

We used the *drop1()* function in RStudio to remove each term from the model, one at a time, and test the changes in the model's fit. The results showed that variables (cm), (csd), and (cr) failed to make any significant contribution and can be easily left out. That finding was confirmed by the stepwise backwards model selection based on Akaike's information criterion (AIC) with the help of the *step()* function in RStudio (Venables & Ripley 2002).

The Model.1 diagnostics with the help of the *influencePlot()* function in RStudio (R Core Team 2013) helped to identify 5 observations as outliers or overly influential data points. After removing those tweets, we fitted to the data **Model.2**, where we left only (mf) mean word frequency, (rl) relative length of tweet, (uwr) unique words ratio, and (ctwr) cross-tweet words representation as the main effects, and we added all possible interactions between those variables.

Comparing Model.1 and Model.2 with the help of the *anova()* function in RStudio proved that Model.2 is significantly better than Model.1: residual deviance decreased by 481.85 (p < 0.001) and AIC decreased from 7081.9 to 7079.2.

Stepwise backward model selection based on AIC suggested that only (mf*rl) mean word frequency by relative length of tweet interaction should be removed. However, given the nature of our variables, we would suggest that a strong multicollinearity exists in the model. Checking for variance inflation factors in Model.2 with the help of the *vif()* function in RStudio (Fox & Weisberg 2018) produced alarming results (Table 8).

Table 8. Variance Inflation Factors in regression model (Model.2)

|  | mf | rl | uwr | ctwr | mf*rl | mf*uwr | mf*ctwr | rl*uwr | rl*ctwr | uwr*ctwr |
|---|---|---|---|---|---|---|---|---|---|---|
| *VIF* | 464.84 | 781.19 | 74.49 | 34.35 | 13.77 | 9.63 | 477.88 | 35.58 | 918.01 | 51.18 |

According to a rule of thumb proposed by Levshina (2015: 160), all values greater than 10 should be regarded as highly linearly related. Having deleted the most highly inflated interactions (mf*ctwr), (rl*ctwr), and (uwr*ctwr), we fitted **Model.3** to the data. Comparing Model.2 and Model.3 with the help of the *anova()* function in RStudio revealed that Model.3 had lost some explanatory power: residual deviance increased by 388.3 ($p < 0.001$). However, multicollinearity was greatly reduced, and goodness-of-fit statistics slightly improved: pseudo-R2 = 0.37 (McFadden), 0.40 (Cox and Snell), 0.53 (Nagelkerke). Stepwise backward model selection based on AIC suggested that interaction (mf*uwr) should be deleted, which was done in **Model.4**. The residual difference increased by 0.97; however, that was not significant ($p = 0.33$).

As the interaction (mf*rl) remained inflated (*vif*-score = 12.39), though significant, we built **Model.5** without taking it into account. Using the *anova()* function in RStudio showed that the residual deviance in Model.5 slightly increased in comparison with Model.4 (21.64, $p < 0.001$). As the difference was marginal, we nevertheless preferred Model.5 to Model.4 because it completely solved the problem of multicollinearity, thus making the results more reliable.

Figure 4. Predicted probabilities of correct tweet classification (Model.5)

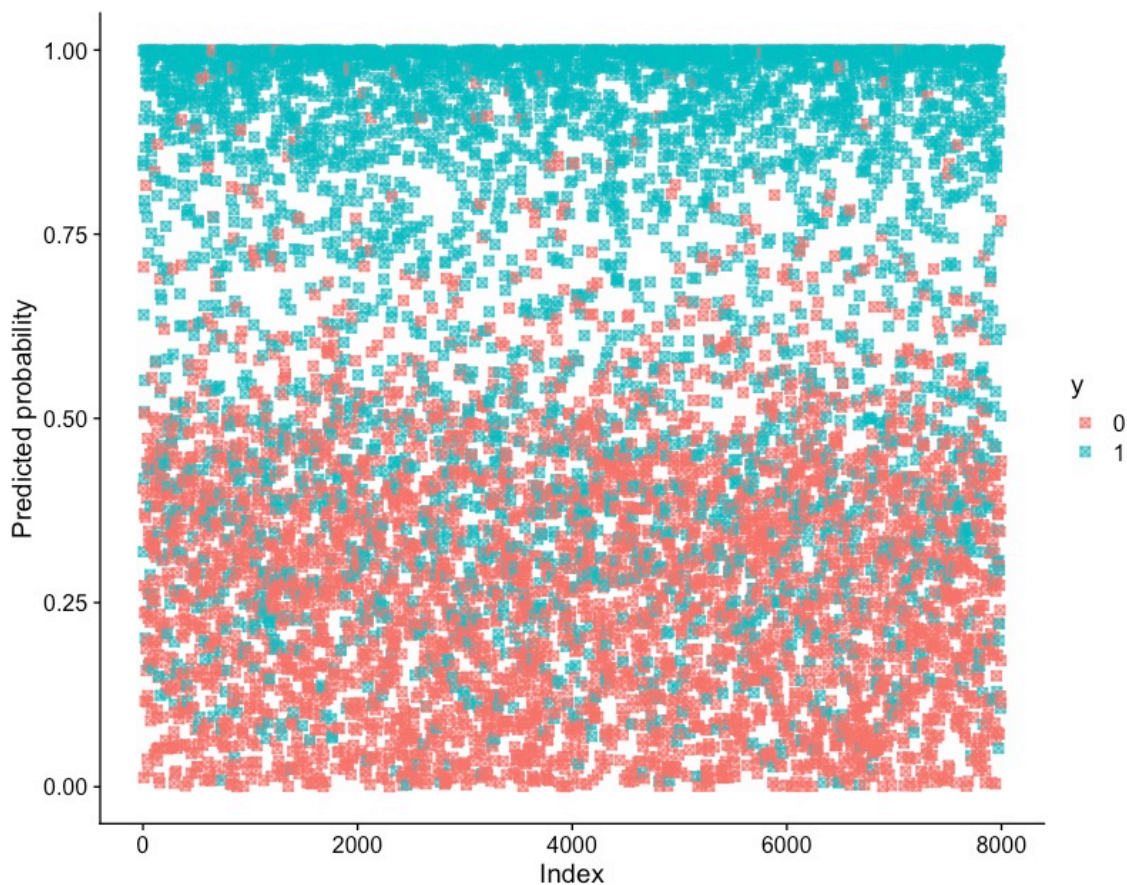

In general, Model.5 did a fairly good job as proved by the regression results presented in Figure 4. The main effects and one interaction of the Model.5 are visualised in Figure 5. The model shows a strong negative effect of the mean word frequency and a strong positive effect of cross-tweet words representation; however, the interaction of relative length and unique word ratio is not as easy to interpret. The interacting variables no longer represent the global change in the classification results with every increase in them; rather, such a change when the interacting term is at its reference level, i.e., it is equal to 0 (Aiken & West 1991).

Figure 5. Effect plot of Model.5

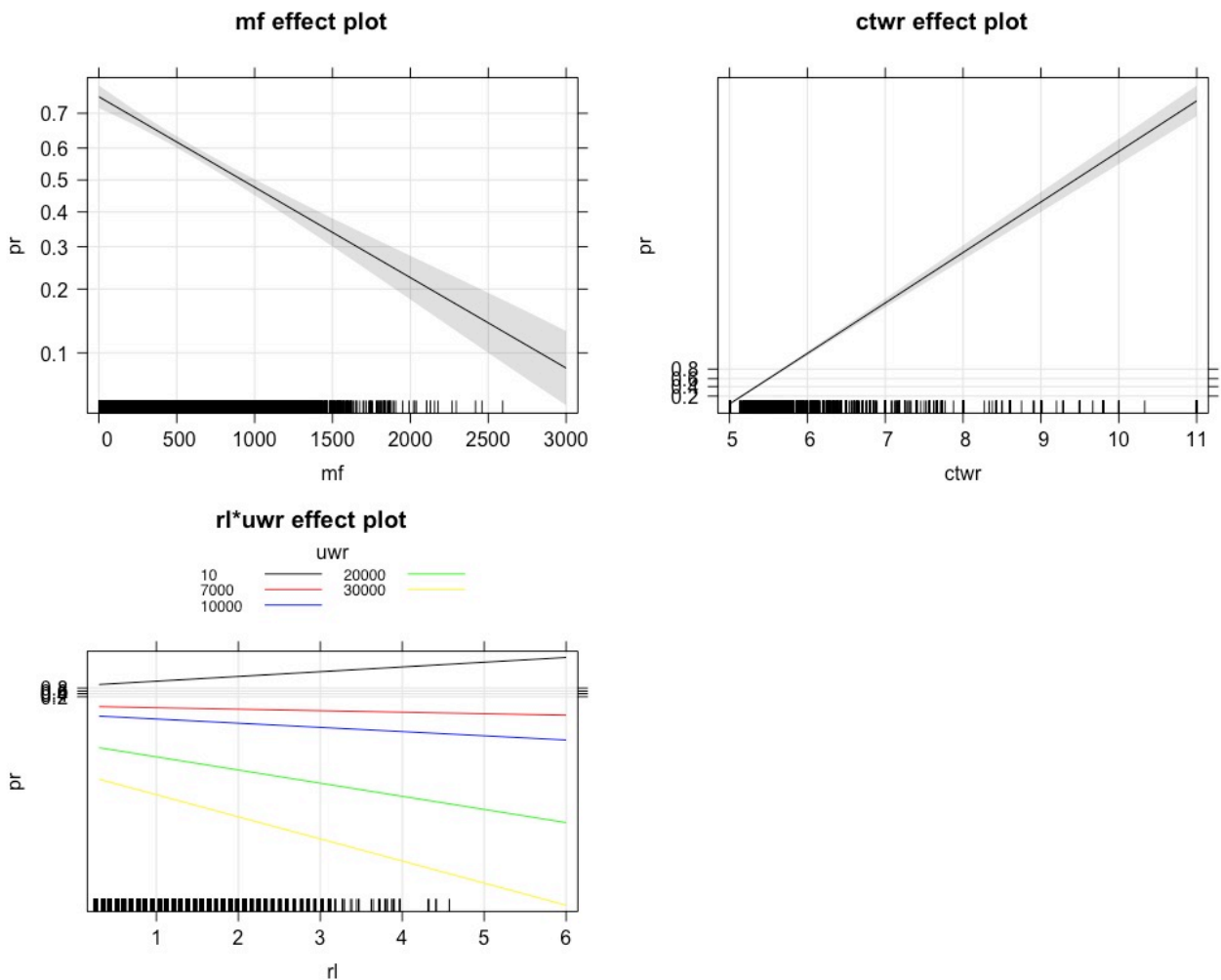

Of course, given the way in which our variables are scaled, 0 is not a particularly meaningful value since it never actually appears in the data: there are no tweets consisting of 0 words or of words with index 0. The effects can therefore be made more interpretable by centring the variables around the mean because when the variables are centred, their reference level corresponds not to meaningless 0 but rather to a perfectly clear average value.

Having done this with the help of the *normalize()* function in RStudio, we obtained the following results (Table 9). The log-likelihood ratio test statistic of Model.5 is 4095.4 with the p-value < 0.0001. Thus, the null hypothesis that the deviance of the model does not differ from the deviance of a model without any predictors can be rejected. Of all the predictors, none failed to cross the threshold of statistical significance (p < 0.0001).

Table 9. Coefficients of regression model (Model.5)

|  | B (SE) | 95 % CI for odds ratio | | |
| --- | --- | --- | --- | --- |
|  |  | Lower | Odds ratio | Upper |
| Constant | 0.38*** (0.03) | 1.37 | 1.47 | 1.58 |
| (mf) mean word frequency | -0.38*** (0.03) | 0.63 | 0.67 | 0.72 |
| (rl) relative length of tweet | 0.41*** (0.03) | 1.40 | 1.51 | 1.63 |
| (uwr) unique words ratio | -3.12*** (0.09) | 0.03 | 0.04 | 0.05 |
| (ctwr) cross-tweet words representation | 4.14*** (0.11) | 50.62 | 63.06 | 79.18 |
| (rl*uwr) | -0.39*** (0.05) | 0.6 | 0.67 | 0.74 |

Note: LR chi2(5) = 4095.4, p < 0.0001. Pseudo-R2 = 0.36 (McFadden), 0.40 (Cox and Snell), 0.53 (Nagelkerke). Significance codes: ***—p < 0.0001.

Overall, we observe both positive and negative main effects that can be interpreted as follows:

(1) the odds of the right versus wrong tweet classification become 0.67 times lower with every increase in mean word frequency,
(2) the odds of the right versus wrong tweet classification become 1.51 times higher with every increase in relative length of tweet when the unique word ratio of the tweet is average,
(3) the odds of the right versus wrong tweet classification become 0.04 times lower with every increase in unique word ratio when the relative length of the tweet is average,
(4) the odds of right versus wrong tweet classification become 63.06 times higher with every increase in cross-tweet words representation,
(5) as the unique words ratio increases, the odds of the right versus wrong tweet classification decrease with every increase in the relative length of the tweet. (Figure 6).

Summing the above, the tweets that are more easily classified share the following characteristics:

(1) include words that are not very frequent,
(2) are longer than average and at the same time do not have very rare words,
(3) contain more words with limited distribution than omnipresent words.

Our last step was to test our regression model on new data. A total of 1,350 observations were randomly selected from the bulk of the *right_tweets* subset that was not used during the training and merged with the remaining 1,350 observations from the *wrong_tweets* subset. Model.5 was fitted to the new data with the help of the *predict()* function in RStudio. To compare the output with the initial class-labels, we replaced all values greater than 0.5 with 1 and all values smaller than 0.5 with 0, thus turning probabilities into categories. The results were evaluated against the levels of the response variable (pr). The accuracy of the prediction turned out to be 85.89 %, which is a fairly decent score showing that our regression model is not overfitted and generalises well.

Next, we return to our neural network and analyse whether there is any difference in the classification of tweets written by trolls and by congressmen. The numbers are as follows:

(1) among the *right_tweets* subset, there are 29,387 troll tweets (53.77 %) and 25,265 tweets by congressmen (46.22 %);
(2) among the *wrong_tweets* subset, there are 2,266 troll tweets (42.34 %) and 3,086 tweets by congressmen (57.66 %).

The results are highly significant, as proved by the Pearson's Chi-squared test with Yates' continuity correction: X-squared = 255.14, df = 1, p-value < 2.2e-16.

Figure 7 shows that the trolls' tweets are overrepresented in the *right_tweets* and underrepresented in the *wrong_tweets* and vice-versa for the congressmen's tweets. We can conclude that there is something intrinsic in troll writing that makes it subject to better labelling (which is, of course, a welcome result), and, given the coefficients of Model.5 and our interpretation of them, we would argue that it is a skewed distribution of at least one or, more likely, several linguistic features.

Figure 7. Mosaic plot of classification results
for troll's tweets and congressmen's tweets

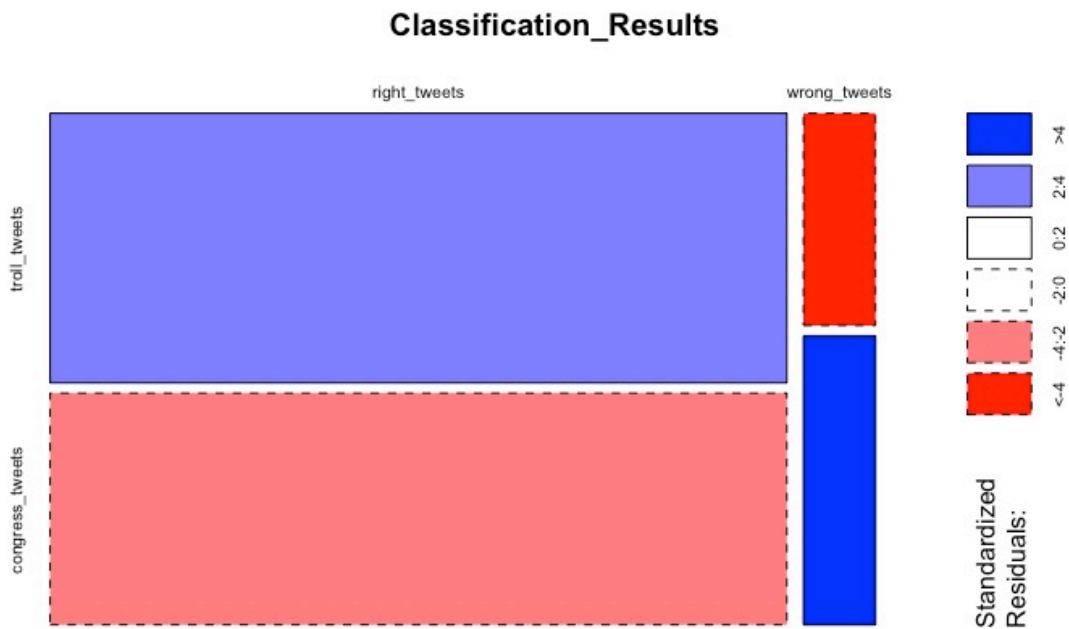

What does being a paid internet troll mean in terms of sociolinguistic behaviour? In essence, it is an imitation game. A troll wants to achieve their goal without being identified as trying to achieve it. In other words, their language attitudes must be predefined and moulded by a combination of two factors: the first one is speaking with a purpose; the second one is trying to mask the purpose of speaking. The different specific weight of these factors and their orthogonal nature necessarily result in the skewed language distribution that we mention above.

The brief summary of possible linguistic features that might be affected by this tendency is provided in Table 10. In this paper, we will briefly consider only two first interconnected features, specifically, the distribution of topics and the vocabulary associated with those topics.

Table 10. Language features that may be anomalously distributed in troll's tweets

|  | Unmarked features | Marked features |
|---|---|---|
| *thematic monotony, recurrent topics* | targeted topics | random new topics |
| *skewed vocabulary* | signal words | random new words |
| *impersonality, lack of empathy* | lack of personal pronouns and determiners | occasional personal pronouns and determiners |
| *repetitiveness* | the same ngrams | occasional new ngrams |
| *exaggeration* | no hedge markers, lots of quantifiers, superlatives | occasional hedge markers, comparatives |
| *future-orientedness* | imperatives, deontic modality, real conditionals | epistemic modality, unreal conditionals |
| *proper nouns* | abundant use of proper nouns | little or no proper nouns at all |
| *polarisation of attitudes* | strong language: damnation and appraisal | soft language: paraphrases and euphemisms |
| *hypertextuality, one-goal-orientedness* | lots of demonstratives, lack of conjunctions | occasional conjunctions |
| *prefabs* | same constructions, same words | same constructions, new words |

**What does it mean to be troll-like: topic modelling**

For topic modelling, we used the implementation of a latent Dirichlet allocation (LDA) algorithm (Blei et al. 2003) in the Gensim package, which is well-known for its efficiency in text classification tasks (Chen et al. 2016). We split our sample of 300,000 tweets into tweets written by trolls and by congressmen and trained multiple LDA models with the number of topics ranging from 3 to 25 for both of the subsets. Then, the CV coherence measure (Röder et al. 2015) was obtained for each topic, and the mean value of all topic coherence measures was calculated for each model.

Table 11. $C_V$ coherence measures for troll's and congressmen's tweets

| Number of topics | Trolls: coherence scores | Congressmen: coherence scores |
|---|---|---|
| 3 | 0.2638 | 0.3495 |
| 5 | 0.2991 | 0.3607 |
| 7 | 0.2788 | 0.4610 |
| 9 | 0.2859 | 0.4439 |
| 11 | 0.3147 | 0.4613 |
| 13 | 0.3472 | 0.4644 |
| 15 | 0.3335 | 0.4913 |
| 17 | **0.3503** | **0.5135** |
| 19 | 0.3338 | 0.4945 |
| 21 | 0.3292 | 0.4880 |
| 23 | 0.3190 | 0.4936 |
| **25** | 0.3248 | 0.4809 |

The results are provided in Table 11. Since the coherence score, which ranges from 0 to 1, is the average of the relative distances between words within a topic, and since all hyperparameters of our LDA models with the sole exception of the number of topics were the same, we would argue that lower coherence scores for topics in the troll tweets serve as evidence of their anomalous lexical distribution. This distribution, as we will show, is in itself a result of using different topics as simple proxies for getting through a very limited number of signals.

First, it is illuminating to observe what the picture will look like if we suppose that all troll tweets are dedicated to just one single topic as are the tweets by congressmen. What words will the LDA models show to be most prominent for this mega topic in either case?

The top 10 words for trolls are as follows:

> 0.009*"trump" + 0.005*"get" + 0.005*"say" + 0.005*"new" + 0.005*"police" + 0.004*"man" + 0.003*"people" + 0.003*"make" + 0.003*"go" + 0.003*"workout"

Their counterparts for congressmen are as follows:

> 0.007*"today" + 0.006*"work" + 0.005*"trump" + 0.005*"thank" + 0.005*"make" + 0.004*"people" + 0.004*"need" + 0.004*"bill" + 0.004*"vote" + 0.004*"help"

We can observe that the major difference is not in the fact that different words have the highest scores. That is what one would expect given the bizarreness of the whole procedure of one-topic-for-all-tweets modelling. What one would not expect is that the distance in coherence scores between the most prominent and second most prominent words is 4 times larger in trolls' tweets compared to congressmen's tweets: 0.004 and 0.001, respectively. This finding means that the word *Trump* tends to be omnipresent in the tweets of trolls, showing up even in the contexts to which it does not naturally belong.

Moreover, if we now obtain the coherence scores of our single topics for their respective corpora, we will obtain **0.3481** for trolls and **0.0729** for congressmen. While the latter value is as marginal as needed to reflect the sheer impossibility of brute-forcing thousands of tweets written by different persons over a considerable time span into just one topic, the former value, in fact, does not deviate much from the coherence scores of models with 5, 19, and even 25 topics trained on the same data.

To confirm and visualise this trend, we plotted mean coherence scores obtained by LDA modelling for both trolls' and congressmen's tweets with a number of topics ranging from 1 to 50. The results (Figures 8 and 9, respectively) are quite remarkable. Thus, the only model where the trolls' tweets surpass the congressmen's tweets in terms of coherence measure is the model with 1 topic. In addition, since this topic as a sequence of top words associated with it is completely unintelligible from a human perspective, we must conclude that the topic modelling of troll writing should take into account its three-level structure that it is very different from that of 'genuine' speech (Figure 10).

We think that troll writing as a linguistic phenomenon is characterised by the omnipresence of a target message (or a small cluster of such messages) underlying each and every concrete topic, however great their range may be. The exceptions to this rule are minor and irregular, motivated by the aforementioned urge to hide the true purpose of speaking. The target messages themselves are nothing more than arrays of several words, which we have called signal or target words that shift from one context to another.

Figure 8. Dynamics of coherence scores with number of topics rising from 1 to 50 in trolls' tweets

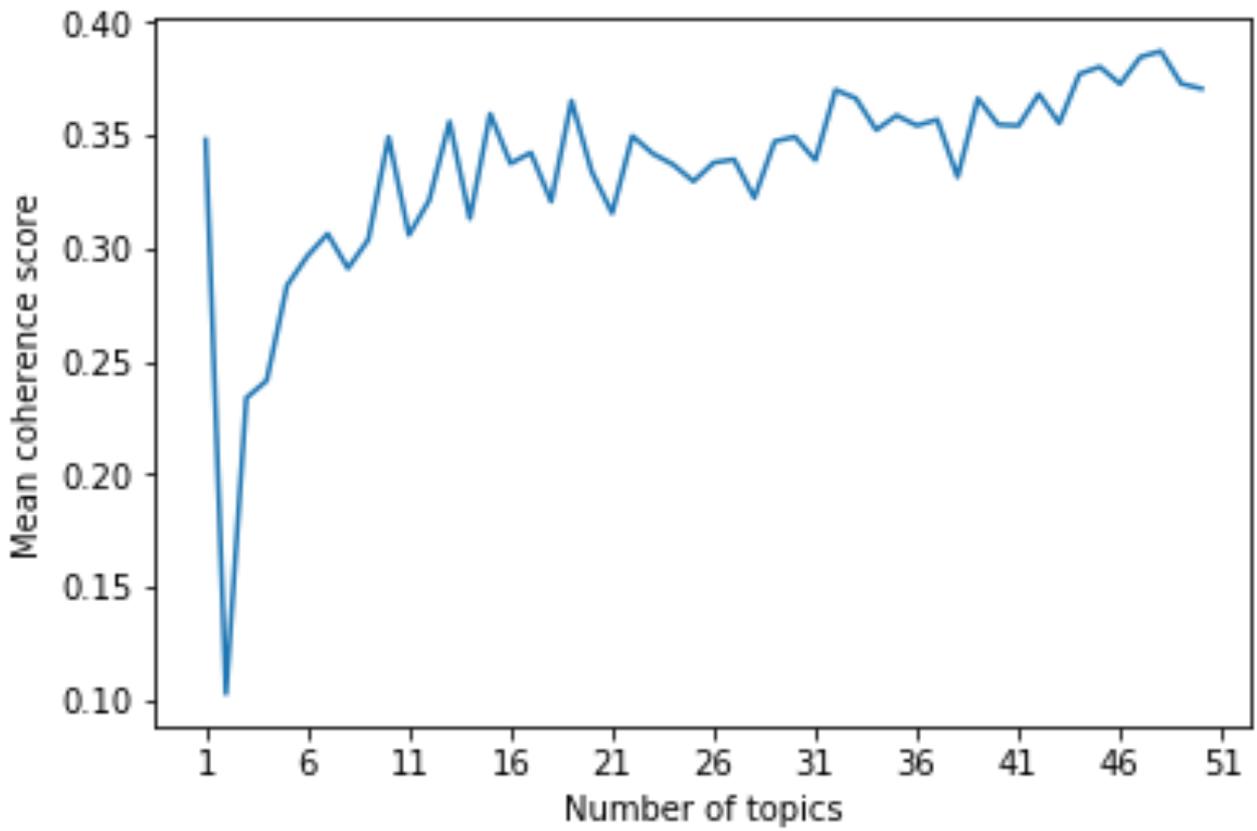

Figure 9. Dynamics of coherence scores with number of topics rising from 1 to 50 in congressmen's tweets

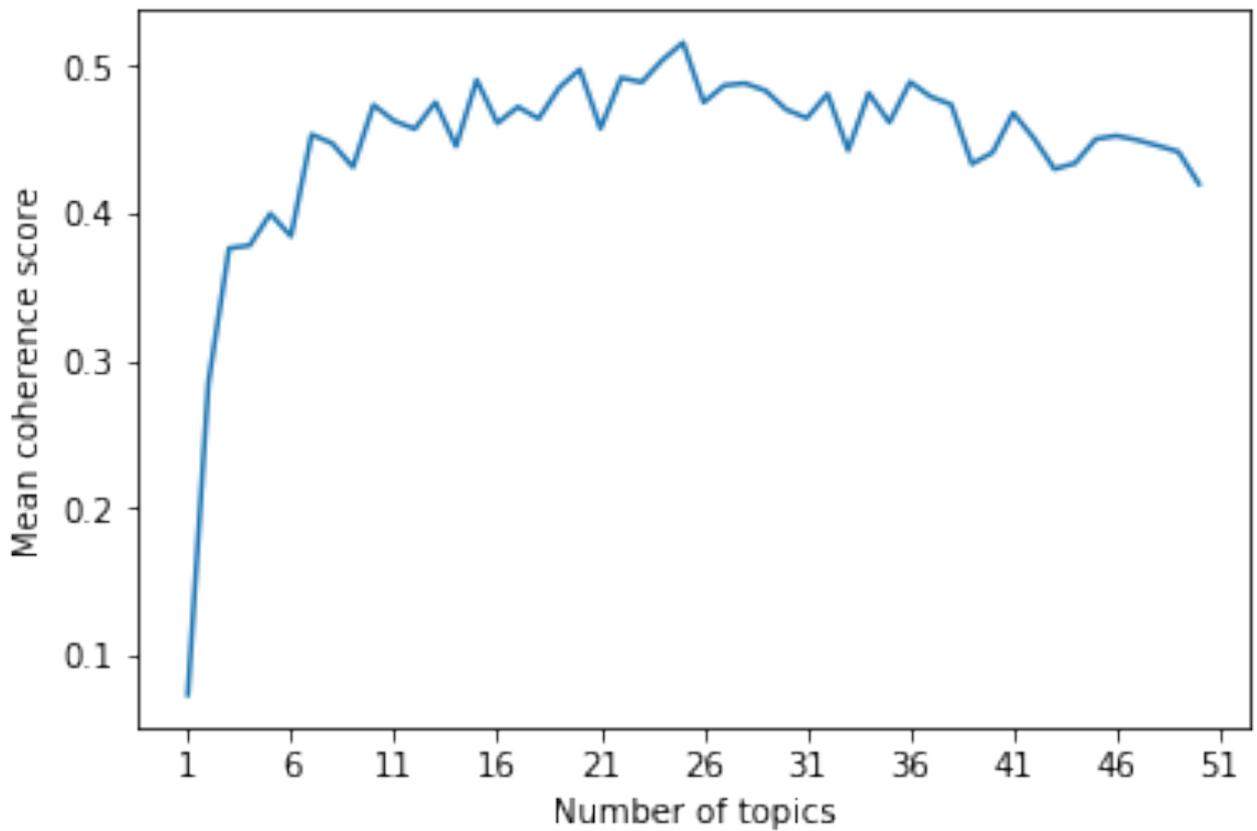

Figure 10. Schema of the three-level structure of troll writing

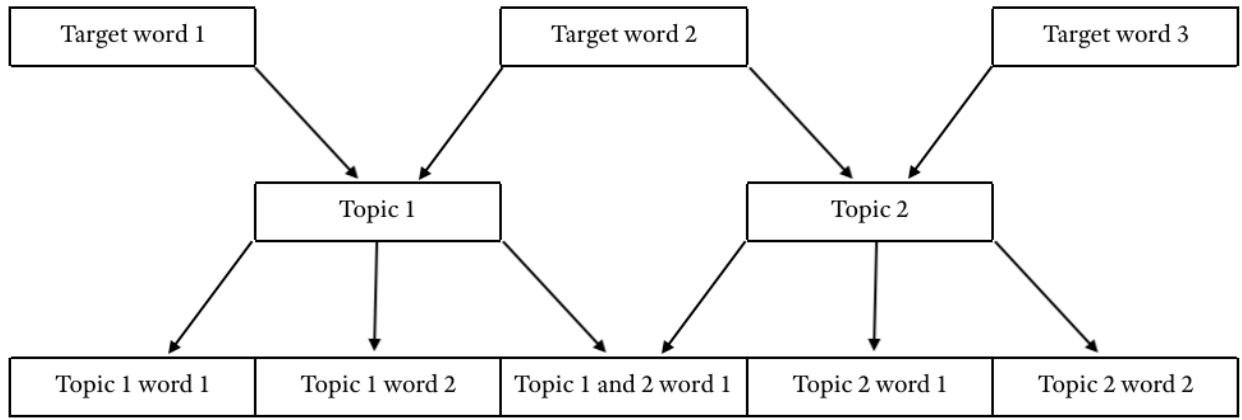

Another observation that supports this hypothesis is based on a comparison of the within-topic coherence scores of the first and second top words associated with a particular topic. To compare these values, we chose the LDA models with the highest overall coherence scores and obtained for each of their 17 topics the 10 most characteristic words. The results of subtracting the score of the second top word from the score of the first one for each topic are presented in Table 12; the standard deviation and mean are shown at the bottom of the table. Arrays of the top words are provided for the topics with the greatest distance value.

Table 12. Distances in coherence measures between the 1st and 2nd top words for different topics

|  | Trolls | | Congressmen | |
| --- | --- | --- | --- | --- |
| Topics | Top words | Distance | Top words | Distance |
| 1 |  | 0.008 |  | 0.002 |
| 2 |  | 0 |  | 0.002 |
| 3 |  | 0.003 |  | 0.012 |
| 4 | 0.046*"rt" + 0.026*"trump" + 0.015*"anti" + 0.013*"russia" + 0.010*"racist" + 0.010*"call" + 0.010*"potus" + 0.008*"north" + 0.008*"us" + 0.008*"say" | **0.020** |  | 0.002 |
| 5 |  | 0.002 |  | 0.006 |
| 6 |  | 0.019 |  | 0.001 |
| 7 |  | 0.003 | 0.048*"tax" + 0.022*"job" + 0.022*"cut" + 0.014*"workers" + 0.012*"pay" + 0.012*"work" + 0.012*"families" + 0.011*"economy" + 0.010*"americans" + 0.010*"american" | **0.026** |
| 8 |  | 0.004 |  | 0.005 |
| 9 | | 0.009 | | 0.007 |
| 10 | 0.059*"police" + 0.037*"shoot" + 0.020*"man" + 0.018*"arrest" + 0.017*"suspect" + 0.017*"fire" + 0.016*"crash" + 0.016*"dead" + 0.015*"kill" + 0.014*"officer" | **0.022** | | 0.005 |
| 11 | | 0.013 | | 0.004 |
| 12 | | 0.001 | | 0.003 |
| 13 | | 0.012 | 0.044*"make" + 0.019*"need" + 0.017*"sure" + 0.016*"every" + 0.015*"keep" + 0.014*"future" + 0.013*"work" + 0.013*"protect" + 0.012*"ensure" + 0.010*"safe" | **0.025** |
| 14 | 0.083*"trump" + 0.028*"obama" + 0.021*"president" + 0.021*"news" + 0.016*"donald" + 0.013*"say" + 0.011*"election" + 0.011*"get" + 0.010*"house" + 0.009*"media" | **0.055** | | 0.005 |
| 15 | | 0.007 | | 0.004 |
| 16 | | 0.005 | | 0 |
| 17 | | 0.011 | 0.049*"trump" + 0.022*"president" + 0.017*"administration" + 0.010*"attack" + 0.007*"north" + 0.006*"nuclear" + 0.006*"war" + 0.006*"de" + 0.006*"call" + 0.006*"korea" | **0.027** |
| | | SD: 0.011 Mean: 0.012 | | SD: 0.008 Mean: 0.008 |

The results show that, on average, in trolls' tweets, the first top word is distinguished from all other words of this topic by a much greater coherence score than in the congressmen's tweets. Even if we leave out the most prominent value of 0.055 in topic 14, the picture will not change in principle: the median values of the distances in the trolls' and the congressmen's tweets are 0.008 and 0.005, respectively.

What can be the reason behind this result? Suppose that one has to write a great number of messages using the word *Trump* in most of them. Of course, it is not possible to simply repeat the same tweet all the time because that will lead to the disclosure of the presence of a troll. Hence, it is necessary to put the target word in a variety of different contexts, including those where it may seem unusual for other speakers. This, in turn, has consequences for the signal word's lexical compatibility: its distribution largely increases, its neighbours become more numerous, and the cooccurrence links between it and other words become unnaturally strengthened.

As a $C_V$ coherence measure is nothing more than the cosine similarity between a target word represented by a vector of its normalised pointwise mutual information (Bouma 2009) with every other word in the top-N topic list, we can predict that the aforementioned distributional anomalies will be reflected in the vector space of the whole corpus.

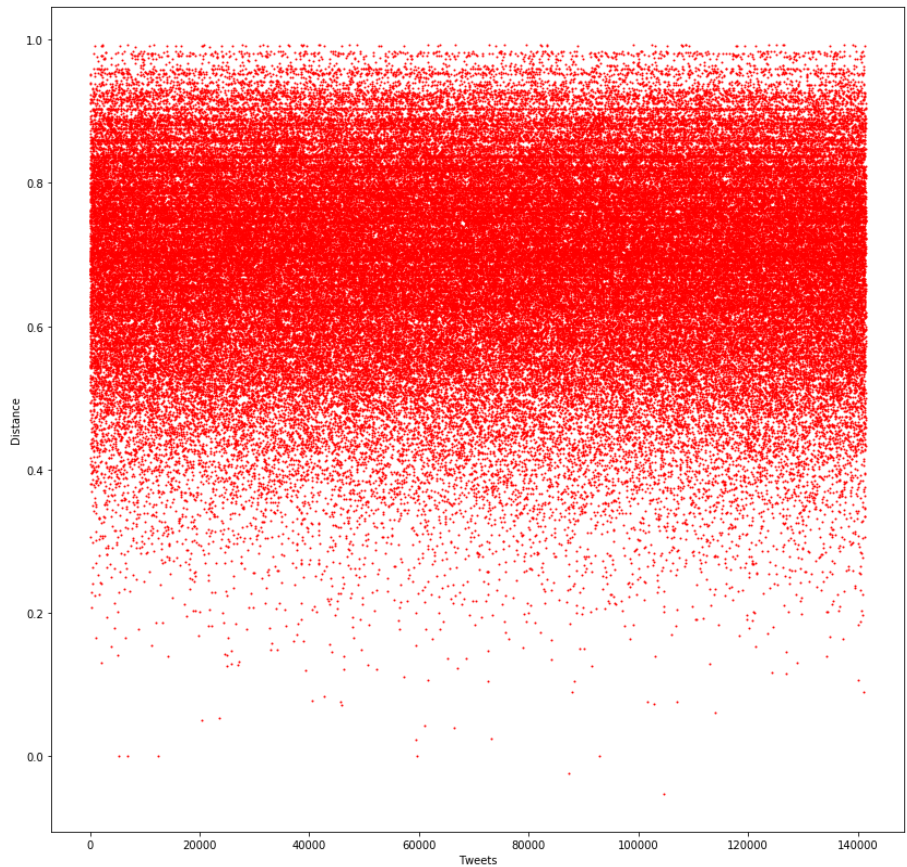

Figure 11. Maximal cosine similarities in congressmen's tweets (SD = 0.13)

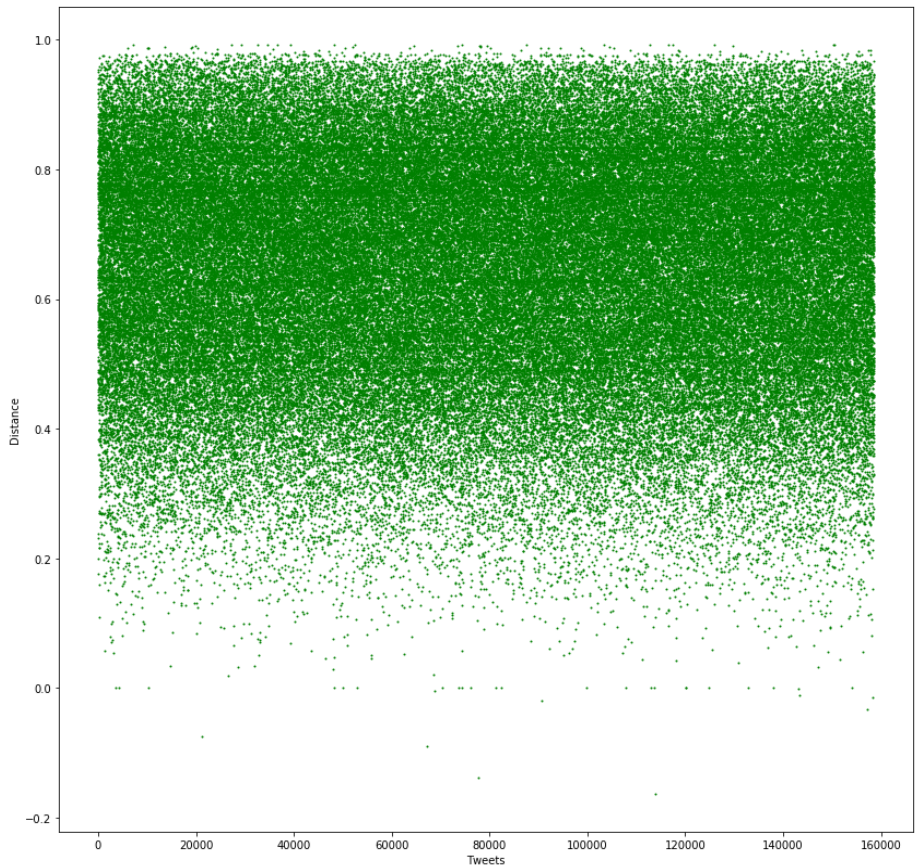

Figure 12. Maximal cosine similarities in trolls' tweets (SD = 0.17)

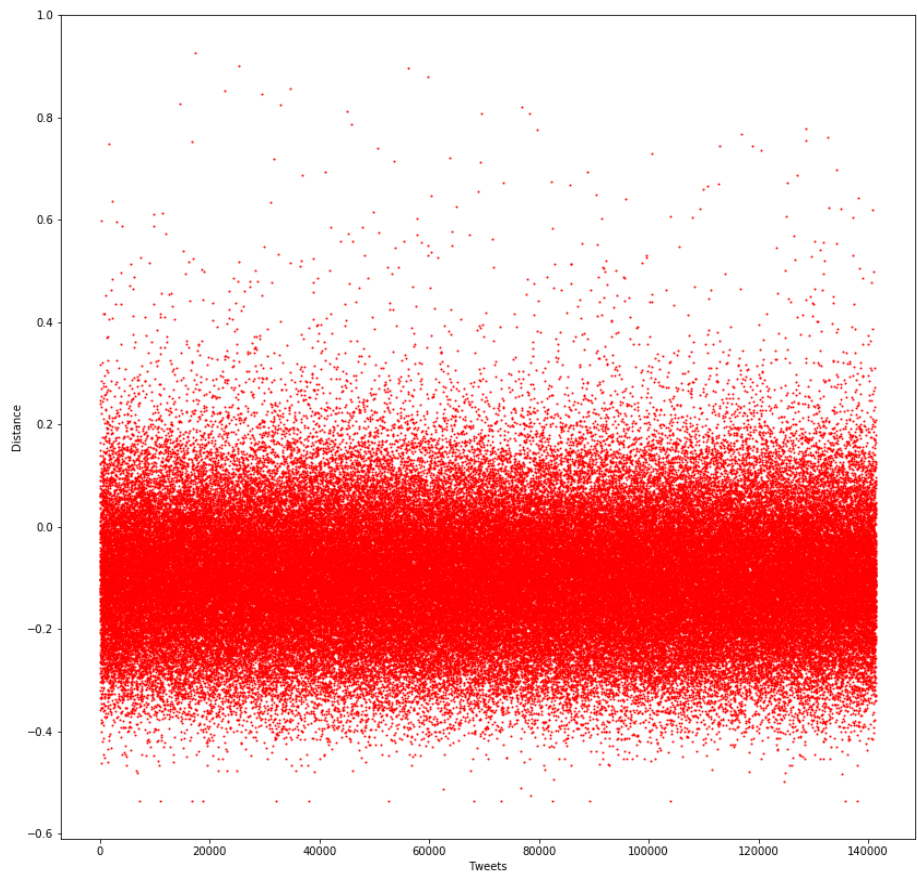

Figure 13. Minimal cosine similarities in congressmen's tweets (SD = 0.12)

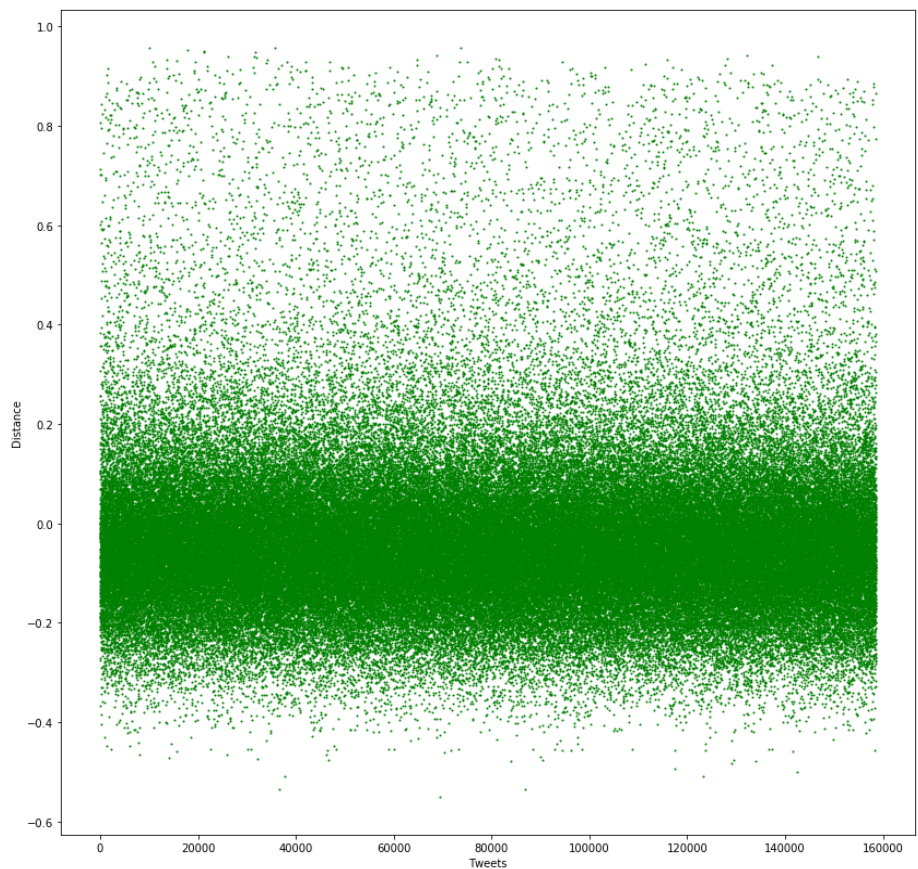

Figure 14. Minimal cosine similarities in trolls' tweets (SD = 0.16)

To test this assumption, we trained a new Word2Vec model on our sample of 300,000 tweets of both trolls and congressmen. The hyperparameters of the model were as follows: the size of the context window equals to 3, the number of iterations equals to 5, the minimal occurrence threshold equals to 1. Thus, every word in the sample was analysed, not only the most frequent ones. Having built this model, we obtained the maximal and minimal cosine similarity values for each tweet and plotted the results (Figures 11–14, respectively). The pattern is, in essence, the same: the dispersion of cosine distances, both maximal and minimal, is greater in the trolls' tweets than in the congressmen's tweets. It suggests that our prediction was accurate.

Two independent one-tailed t-tests with Welch's correction were performed to compare the means of the maximal and minimal cosine distances in 2,000 randomly selected congressmen's and trolls' tweets. With maximal values, on average, the congressmen's tweets produced significantly greater cosine similarities (M = 0.70, SE = 0.004) than the trolls' tweets (M = 0.64, SE = 0.004), t(1903.2) = 8.0256, p < 0.0001. Conversely, with minimal values, the congressmen's tweets are characterised by significantly lower cosine similarities (M = -0.10, SE = 0.003) than the trolls' tweets (M = -0.03, SE = 0.005), t(1861.1) = -10.926, p < 0.0001. Respective bar plots of the sample means with 95 % confidence intervals are provided in Figures 15 and 16.[5]

Figure 15. Box plots of maximal cosine distances in trolls' and congressmen's tweets with 95 % confidence intervals (word count threshold = 1)

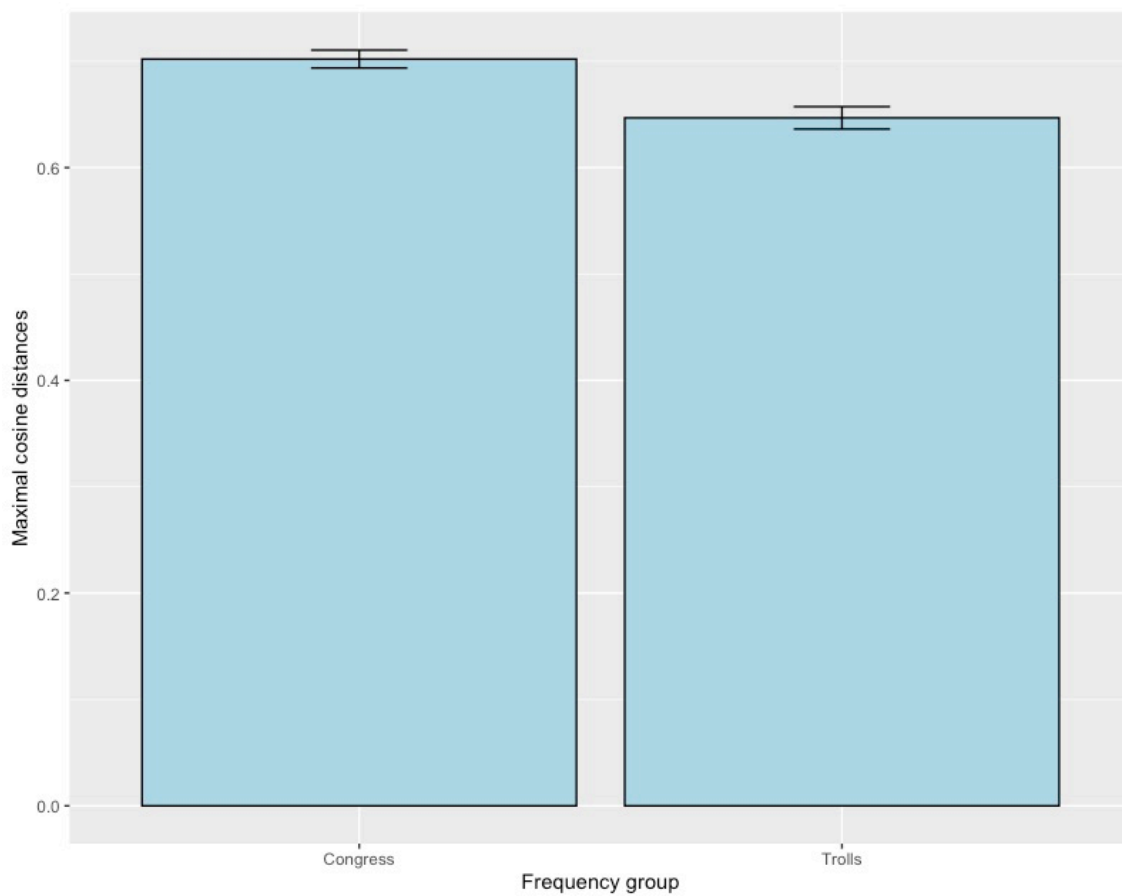

---

[5] The same pattern, though with less pronounced difference between minimal cosine distances, holds if we take into account only distribution of words that appear at least 5 times in the corpus: with maximal values, on average, the congressmen's tweets produced significantly greater cosine similarities (M = 0.53, SE = 0.004) than the trolls' tweets (M = 0.38, SE = 0.005), t(1972) = 21.243, p < 0.0001. Conversely, with minimal values, the congressmen's tweets are characterised by significantly lower cosine similarities (M = -0.28, SE = 0.004) than the trolls' tweets (M = -0.22, SE = 0.004), t(1972) = -9.595, p < 0.0001.

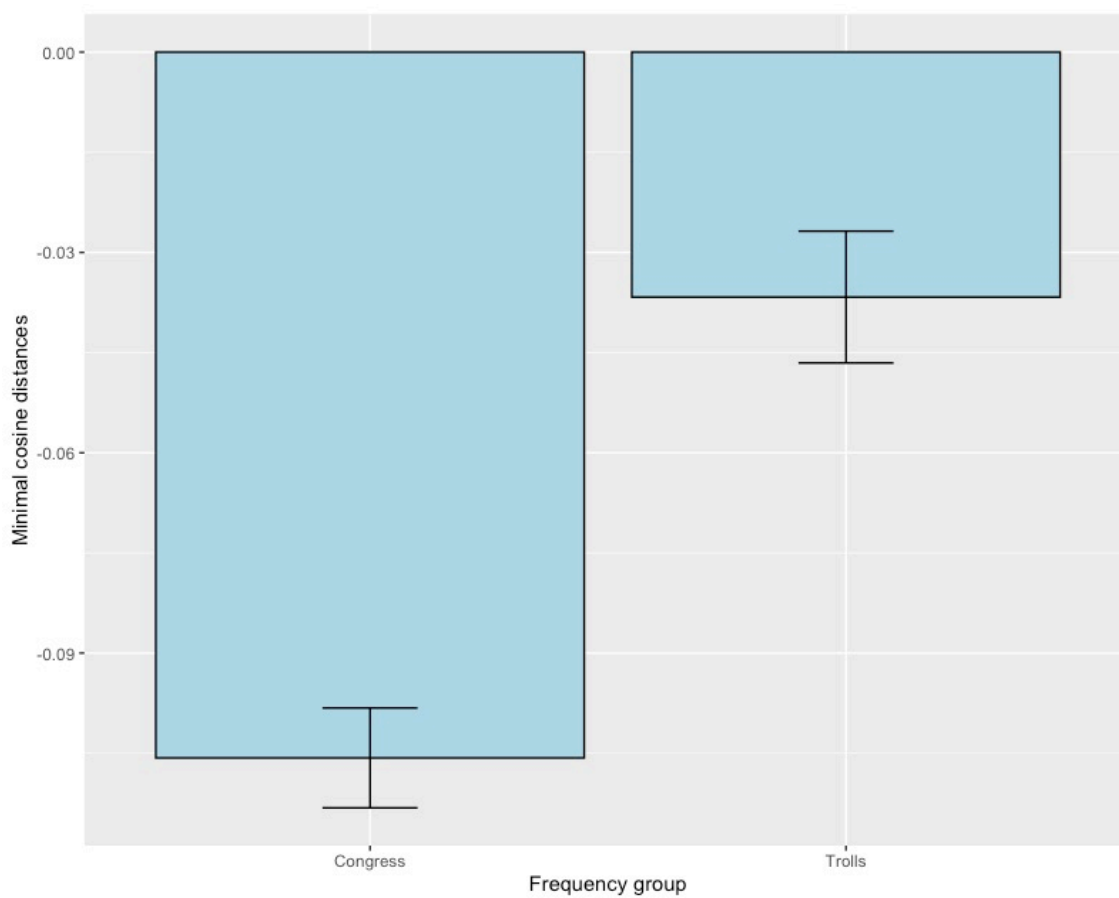

Figure 16. Box plots of minimal cosine distances in trolls' and congressmen's tweets with 95 % confidence intervals (word count threshold = 1)

We think that such a distribution can only be explained by the fact that certain signal words, which are important for the true purpose of trolls' tweets, are forced upon multiple contexts, thus distorting all cooccurrence measures in the corpus.

Let us consider the situation with 'genuine' tweets, as we consider the congressmen's tweets to be. High maximal and low minimal cosine similarity values mean that, on average, in each tweet there tends to be at least one pair of words that are frequently used together in the corpus and at least one pair of words that are rarely used together. In contrast, less extreme maximal and minimal cosine similarity values, which are observable in the trolls' tweets, mean that, on average, all words of that tweet appear together throughout the corpus quite a number of times, although not always in the same combination. As J. R. Firth's famous maxima goes, '[y]ou shall know a word by the company it keeps' (Firth 1957: 11). Thus, between the words in the trolls' tweets, there are no close friends nor bitter enemies: only good acquaintances.

One important deduction that can be made from these results is that each troll's tweet in terms of its words' cooccurrence measures is likely to be more isomorphic to the whole corpus, i.e., to act as the corpus's better representative than any 'genuine' tweet. That perfectly explains the anomalous one-topic-for-all LDA model's coherence score that was mentioned above.

**Conclusion**

The current study yielded a number of important findings. We managed to build a neural network that achieved an accuracy score of 91 % in classifying troll and 'genuine' tweets. By means of regression analysis, we identified a number of features that make a tweet more susceptible to correct labelling and found that they are inherently present in trolls' tweets as a special type of discourse.

We hypothesised that those features are grounded in the sociolinguistic limitations of troll writing, which can be best described as a combination of two factors: speaking with a purpose and trying to mask the purpose of speaking. Next, we contended that the orthogonal nature of these factors must necessarily result in the skewed distribution of many different language parameters of trolls' messages. Having chosen as an example distribution of the topics and vocabulary associated with those topics, we showed some very pronounced distributional anomalies, thus confirming our prediction.

It is natural to assume that very similar anomalies will be identified on many other levels of trolls' writing that for now were left beyond the scope of the study. Among them, we could name personal pronouns and demonstratives, different types of ngrams, hedge markers, quantifiers, superlatives, imperatives, deontic modals, real conditionals, conjunctions, and different types of constructions.

We have no doubt that in the future, the task of identifying trolls' internet messages will be solved, and such posts will be filtered out as easily and successfully as spam emails are today. The importance of this work for modern society can hardly be overestimated.

**Data sources:**

Collection of Russian Troll Tweets. Available online at <https://github.com/fivethirtyeight/russian-troll-tweets/>. June 2019.
Collection of Tweets of Congress. Available online at <https://github.com/alexlitel/congresstweets/>. June 2019.
Collection of Trump Tweets. Available online at <https://github.com/mkearney/trumptweets>. June 2019.